\documentclass[twocolumn,twosided,english,nofootinbib,aps]{revtex4-1}

\usepackage{color}
\usepackage{amsmath}
\usepackage{graphicx}
\usepackage{amssymb}
\usepackage{esint}
\usepackage[
hyperfootnotes=true,
unicode=true, 
 bookmarks=true,bookmarksnumbered=false,bookmarksopen=false,
 breaklinks=false,pdfborder={0 0 1},backref=false,colorlinks=true]
 {hyperref}

\makeatletter

\@ifundefined{textcolor}{}
{%
 \definecolor{BLACK}{gray}{0}
 \definecolor{WHITE}{gray}{1}
 \definecolor{RED}{rgb}{1,0,0}
 \definecolor{GREEN}{rgb}{0,1,0}
 \definecolor{BLUE}{rgb}{0,0,1}
 \definecolor{CYAN}{cmyk}{1,0,0,0}
 \definecolor{MAGENTA}{cmyk}{0,1,0,0}
 \definecolor{YELLOW}{cmyk}{0,0,1,0}
 }

\usepackage{hyperref}
\hypersetup{
    colorlinks,%
    citecolor=blue,%
    filecolor=blue,%
    linkcolor=blue,%
    urlcolor=blue
}

\makeatother

\usepackage{enumitem}

\newcommand{\be}{\begin{eqnarray}}
\newcommand{\ee}{\end{eqnarray}}

\begin{document}


\title{
Is Gauge Mediation in the Swampland? 
}

\author{Ioannis Dalianis}
\affiliation{ Department of Physics, University of Athens, \\ 
            University Campus, Zographou 157 84, Greece}

\author{Fotis Farakos}
\affiliation{ Dipartimento di Fisica e Astronomia ``Galileo Galilei'' \\ 
            Universit\`a di Padova, Via Marzolo 8, 35131 Padova, Italy} 
\affiliation{ INFN, Sezione di Padova \\ 
            Via Marzolo 8, 35131 Padova, Italy}            

\author{Alex Kehagias}
\affiliation{ Physics Division, National Technical University of Athens, \\ 
            15780 Zografou Campus, Athens, Greece}


\begin{abstract}

We note that the typical gauge mediation of supersymmetry breaking is in tension with the global limit of the festina lente swampland bound. The alternatives are mediation/breaking schemes that decouple together with gravity, as for example gravity mediation, for which we highlight some basic phenomenological properties. Gauge mediation remains instead a viable mechanism only in models where supersymmetry is restored in the global limit, as for example in no-scale supergravity. 

\end{abstract}

\maketitle

\section{Introduction}

Supersymmetry has seen a variety of motivations throughout the years. 
Originally, in the Volkov--Akulov model \cite{Volkov:1973ix}, 
it was considered as a natural way to keep the neutrino massless due to the underlying non-linear realization. 
Later on, once the linear realization of supersymmetry was constructed \cite{Wess:1974tw}, 
it became clear that such a symmetry could tame radiative corrections, 
and protect the electro-weak scale from ultra-violet (UV) physics \cite{Martin:1997ns,Weinberg:2000cr,Aitchison:2007fn}. 
As an added bonus, 
supersymmetry made gauge coupling unification \cite{Dimopoulos:1981yj} more feasible, 
and also offered a variety of candidates for dark matter particles. 
For the moment, 
no signal of supersymmetry has been observed in colliders, 
therefore the realization of such symmetry in particle physics is still unclear. 
In particular, 
the central problem of supersymmetrizing the Standard Model (SM), 
is that, after all, supersymmetry has to be broken. 
The ambiguity then in the phenomenology comes from the vast freedom in deciding on a mediation scheme. 
This ambiguity leads to a large number of unknown new couplings and fields that enter the model-building procedure.

The advantage of supersymmetry, however, is that it is also motivated by string theory \cite{Weinberg:2000cr,Becker:2006dvp}, 
that is, a theory of quantum gravity. 
Actually, due to the null experimental signals, 
it is the string theory motivation of supersymmetry that carries the day. 
This means that the scale of supersymmetry breaking may be quite high in our world 
(if for instance the source is some type of brane-supersymmetry-breaking \cite{Mourad:2017rrl}), 
and in addition, the mediation of the breaking to the SM may be quite sensitive to the UV completion. 
In this vein, 
one would like to be aware of possible restrictions both on the mediation and the breaking of supersymmetry from string theory/quantum gravity. 
One indirect way of deducing restrictions to the low energy field theory from quantum gravity is the swampland program \cite{Vafa:2005ui,Arkani-Hamed:2006emk,Ooguri:2006in}. 
This program is based on the concept that not all effective field theories can be embedded in quantum gravity, 
and that, we can reliably deduce basic principles dictating what type of theories cannot enjoy such a UV completion 
(for a diverse list of recent reviews see e.g., \cite{Agmon:2022thq,Cribiori:2023gcy,VanRiet:2023pnx}).

The various restrictions are typically bounds on the couplings and the theories that do not satisfy them are said to be in the swampland. 
These restrictions are derived and tested in a plethora of ways. 
One method, 
is to deduce conditions typically in regions of the parameter space where string theory calculations are well under control, 
and then the results are extrapolated to hold generically; 
this last assumption is possibly one of the weakest parts of the swampland program because there is no actual guarantee that such property should hold. 
Another way to deduce swampland bounds is by studying the properties of black holes. 
Indeed this method is very fruitful and has given rise to the weak gravity conjecture \cite{Arkani-Hamed:2006emk}, 
one of the most trusted swampland criteria, 
which has been also studied in asymptotically de Sitter and AdS spaces \cite{Huang:2006hc}. 
A further study of large black holes in de Sitter has given rise to a newer bound, 
referred to as ``festina lente'' (FL) in \cite{Montero:2019ekk}.

What should be stressed about the swampland program is that it refers to effective field theories that are coupled to gravitation. 
Therefore, in the limit where gravity is decoupled one typical expectation is that the swampland bounds should become empty. 
This is not the case for the FL bound however, 
where the constraint becomes singular as $M_P \to \infty$ \cite{Montero:2019ekk}, 
leaving the global limit as an open question. 
In \cite{Montero:2021otb} a version of the FL bound was investigated that allows the decoupling of gravity but still leads to a smooth non-trivial bound in the global limit.

In this work, 
we will discuss exactly how the aforementioned global FL bound (GFL), when taken at face value, 
can help us eliminate large classes of phenomenological models related to supersymmetry. 
In particular, 
we will see that models where the supersymmetry breaking is communicated to supersymmetric versions of the SM (e.g. the MSSM \cite{Martin:1997ns,Weinberg:2000cr,Aitchison:2007fn}) through gauge mediation \cite{Giudice:1998bp} are in tension with the GFL bound.\footnote{Such a difficulty has been anticipated in some works that search directly for gauge mediation within string theory constructions, 
as for example \cite{Dudas:2008qf,deAlwis:2010ud}. It is gratifying to see that the swampland program can give a bottom-up interpretation/verification of such expectation, which in turn gives further evidence in favor of the GFL bound. } 
Notably, models where the mediation is shut down in the global limit evade this bound. 
For instance these can be models of gravity mediation \cite{Hall:1983iz}. 
Particular gauge mediation schemes  of supersymmetry breaking are in a tough spot also from the bounds set by collider physics (see e.g., \cite{Lu:2017oee}) 
therefore the experimental bounds resonate with the consequences of the GFL bound. 
In addition, as we are entering an era where effective field theories of the SM, and beyond the SM, are becoming important, 
it is crucial to have a guide for which higher dimensional operators we should expect to dominate. 
Having a specific mediation scheme favored for the supersymmetry breaking is therefore welcome. 
Conversely, 
applying swampland bounds (or candidate swampland bounds) to phenomenology also provides a valuable test for such criteria.

\section{Swampland, FL and SUSY} 

The FL bound was introduced in \cite{Montero:2019ekk}. 
It applies to an abelian gauge field, either belonging to a U(1) gauge group or to a Cartan of a bigger group. 
The bound is deduced from the decay of large charged, so called Nariai, black holes in de Sitter space. 
The idea is that even though the black hole should decay it should not decay too fast. 
To avoid such fast decay one sets restrictions on the matter content of the theory which, 
in the presence of gravity, can be cast in the form 
\be
\label{FL}
m^4 \gtrsim  g^2 q^2 H^2 M_P^2 \quad , \quad \text{for every charged particle.} 
\ee
Here $m$ is the mass, $g$ is the gauge coupling for an abelian gauging, $q$ is the charge, and $H$ is the Hubble constant. 
Therefore this condition holds on de Sitter in the presence of gravity.

The constraint \eqref{FL} remains non-trivial in the formal global limit, 
that is, 
\be
\label{GD}
\text{gravity decoupling:} \quad M_P \to \infty \,. 
\ee
The question of taking such limit for the FL remains open, 
but it was studied in \cite{Montero:2021otb} where a possible global limit was investigated. 
In search for a smooth behavior under the limit \eqref{GD} the bound is recast in the form 
\be
\label{GFL}
m^4 > g^2 q^2 V \quad , \quad \text{for every charged particle.} 
\ee 
Now $V$ is the vacuum energy of the field theory and because no $M_P$ is explicitly appearing, 
it is speculated that such a bound is worth considering further because it may in the end correspond to the actual correct global limit of the FL bound. 
It was of course readily stressed that the global constraint \eqref{GFL} 
may not really be the final correct version, 
and that refinements may be needed, 
however it passed few basic consistency checks. 
For example, 
the dynamical supersymmetry breaking was studied in \cite{Montero:2021otb}, 
and it was found to satisfy the GFL bound \eqref{GFL}. 
In addition, supersymmetric vacua automatically satisfy the GFL bound.

Let us then take the bound at face value and study it in our physical world applying it directly to the SM. 
Clearly, we can consider the particle content and the masses of the SM particles to be the ones corresponding to the limit \eqref{GD}. 
Therefore we can apply the condition \eqref{GFL} on the electron, which is charged under the U(1) of electromagnetism. 
We have that 
\be
m_{el} \simeq 0.5 \times 10^{-3} {\rm GeV} \quad , \quad (gq)_{el}^2 \simeq 4 \pi \alpha \sim 0.1 \,, 
\ee
which, once inserted in \eqref{GFL}, lead to 
\be
\label{ellFL}
10^{-12} ({\rm GeV})^4 > 10^{-2} V \quad \to \quad V < 10^{-10} ({\rm GeV})^4 \,. 
\ee
Such condition is of course toothless in a field theory because there is no meaning in defining an absolute zero vacuum energy.

This ambiguity is indeed recognized in \cite{Montero:2021otb}, 
and it is then proposed that maybe the condition should be considered within supersymmetric field theories; 
there the vacuum has a well-defined energy value and zero vacuum energy corresponds to a supersymmetric vacuum. 
Indeed, 
in a supersymmetric theory the couplings are controlled by the K\"ahler potential $K (A^i, \overline A^j)$, 
which depends on the scalars of the supersymmetric theory, 
the superpotential $W(A^i)$ and the type of gauging that one performs. 
The potential energy is 
\be
\label{SP}
V = (K_{i\overline j})^{-1} W_i \overline W_{\overline j} + \frac12 (f_{ab})^{-1} D^{(a)} D^{(b)} \,, 
\ee 
where $D^{(a)}$ are field-dependent functions that depend on the gauging, 
$(a)$ is an index that runs over the generators of the gauge group, 
and $f_{ab}$ is the gauge kinetic function which typically is $\delta_{ab}/g_{(a)}^2$. 
The reader can find the details in \cite{Wess:1992cp}. 
The condition that supersymmetry is preserved on the vacuum is that $\langle W_i \rangle= 0 = \langle D^{(a)} \rangle$, 
and therefore the vacuum energy has to vanish on a supersymmetric vacuum; 
the inverse also holds when there are no negative norm states in the system. 
Therefore the vacuum energy can be calibrated in this way to vanish when supersymmetry is preserved. 
It is actually intriguing that the GFL bound makes sense when applied to a supersymmetric field theory. 
Evidently, if one has two split sectors with 
\be
\label{split12}
K = K^{(1)} + K^{(2)} \quad , \quad W= W^{(1)} + W^{(2)} \,, 
\ee
with different fields in sector-1 and sector-2, 
then one should apply the GFL independently to sector-1 and sector-2. 
This makes sense because the two sectors are two completely decoupled field theories.


A point that is worth making concerns the actual interpretation of the bound \eqref{GFL}. 
One way to think of the constraint is that it applies directly to theories with global supersymmetry and that if such constraint is not satisfied then they are for some reason inherently sick. 
This is a {\it strong} version of the GFL and since we are not aware of any complementary derivation that can be performed without the presence of gravity it is difficult to support it. 
Alternatively, 
one can think of the GFL as a bound that actually refers to systems where the coupling to gravitation is made arbitrarily small. 
In this {\it weak} version, 
the bound is interpreted as a limiting procedure of a theory coupled to gravity where $M_P \to \infty$ is a formal limit where the GFL applies. 
The weak version of the GFL therefore is interpreted as a by-product of the full FL bound and means that one does not have to find independent arguments in favor of the GFL directly in a theory with global supersymmetry.

\section{Mediation of SUSY Breaking}

We have seen that it is meaningful to investigate the implications of the GFL bound in systems with global supersymmetry and apply it to each sector independently. 
This will allow us to make an observation related to the mediation of the supersymmetry breaking from the hidden sector to the SM. 
From the properties of the potential \eqref{SP} we know that when supersymmetry is spontaneously broken, 
unavoidably the vacuum energy will have the form 
\be
\label{Vf2}
\langle V \rangle = f^2 > 0 \,, 
\ee
where we can identify 
\be
\sqrt{f} = \text{supersymmetry breaking scale.}
\ee
This property is completely model independent (even if we consider higher derivatives or loop corrections), 
and holds no matter how we break supersymmetry, as long as it is spontaneous, i.e. a goldstino exists. 
Note that adding further (decoupled or not) supersymmetry-breaking sectors simply gives further (possibly distinct) additive positive contributions. 
In other words, in global supersymmetry, 
$\langle V \rangle$ is always greater or equal to the characteristic vacuum energy of each of the supersymmetry breaking sectors. 
For example for a system like \eqref{split12} we would have $\langle V \rangle = \langle V^{(1)} \rangle + \langle V^{(2)} \rangle$.

In a typical phenomenological setup one assumes that we have a supersymmetric version of the SM 
where all particles have a superpartner, 
and a different hidden sector where the breaking happens. 
Then one mediates the breaking of supersymmetry from the hidden sector to the SM. 
This can happen in different ways and it constitutes the mediation of supersymmetry breaking. 
 Assume now that the breaking of supersymmetry is indeed mediated to the SM
with a setup of the form 
\be
\label{GMSB}
K = K^{\rm (SM + BREAK)} \ , \ \ 
W = W^{\rm (SM + BREAK)} \,, 
\ee
where the sectors mix.
Therefore the GFL condition has to hold again for the electron (or any other SM particle). 
From our previous discussion we know that the vacuum energy will be controlled by the supersymmetry breaking \eqref{Vf2}. 
Thus from \eqref{ellFL} we find 
\be
f^2 < 10^{-10} ({\rm GeV})^4 \quad \to \quad \sqrt{f} < 10^{-3} {\rm GeV} \,. 
\ee
We see that the GFL bound is badly violated in such scenario if we assume a realistic  $\sqrt{f} \gtrsim  \, {\rm TeV}$.

Gauge mediation corresponds essentially to a system of the form \eqref{GMSB}, 
therefore it is in tension with the GFL bound. 
The only exception could be a case where $f$ itself goes to zero in the $M_P \to \infty$ limit.

To avoid this conundrum, without restoring supersymmetry, there is a simple condition on the mediation of supersymmetry breaking: 
the mediation has to be {\it gravitational}. 
This means we will have 
\be
\begin{aligned}
K &= K^{\rm (SM)} + K^{\rm (BREAK)} \,, 
\\[0.3cm]  
W &= W^{\rm (SM)} + W^{\rm (BREAK)} \,. 
\end{aligned} 
\ee
Here the supersymmetry breaking is mediated from the ``BREAK'' sector to the ``SM'' sector via $1/M_P$ terms \cite{Hall:1983iz}, 
which vanish in the formal global limit. 
These $1/M_P$ terms are introduced directly from the structure of the supergravity theory \cite{Wess:1992cp,DallAgata:2021uvl}. 
Thus, clearly, in the {\it formal} global limit the GFL can be satisfied, 
but this limit does not correspond to the effective low-energy theory. 
The latter is found after following the rules from gravity mediation (as analyzed e.g. in \cite{Hall:1983iz}).\footnote{Let us highlight once more that there exist different limiting procedures for the decoupling of gravity \cite{deAlwis:2012aa}. 
However, there is only a single limiting procedure that delivers global supersymmetry when $M_P \to \infty$. 
The latter is what we call here {\it formal} global limit and it is the (only) limit where the GFL applies \cite{Montero:2021otb}.} 
In the formal global limit where supersymmetry is preserved in the visible sector the GFL bound is respected. 
This happens because in typical supersymmetric versions of the SM, 
supersymmetry is preserved by excluding the mediation of supersymmetry breaking.

Of course one has to assume that appropriate supersymmetry breaking sectors exist that do 
not violate the FL bound in the global limit. 
This has been certified/checked in \cite{Montero:2021otb} where it is shown that dynamical supersymmetry breaking, 
for example, is compatible with the GFL bound. 
In particular, they discussed the ISS model \cite{Intriligator:2006dd}, which describes a BREAK sector with  a single $U(1)$ gauge field at low energies. The theory features a metastable susy breaking vacuum with $V\sim \Lambda^2 m^2=f^2$, where $m$ is the quark mass scale and $\Lambda$ is the strong coupling scale, and the GFL bound is satisfied since the charged baryons acquire a mass $\sim \Lambda\gg m$.  
In the case that the BREAK sector is characterized by singlet fields (e.g. a nilpotent goldstino \cite{Rocek:1978nb,Casalbuoni:1988xh,DallAgata:2016syy}), 
which is typical in SUGRA mediation schemes, 
the condition $m^4_{} > g^2 q^2 V$ is trivially satisfied. 
Therefore, it seems that the FL bound does not necessarily provide constraints concerning the energy scale of supersymmetry breaking $\sqrt{f}$, nevertheless, as we noticed, 
it has non-trivial implications for the mediation scheme of supersymmetry breaking to the SM. 
Further consequences for non-supersymmetric systems have been discussed for example in \cite{Ban:2022jgm}.

\section{Discussion}
{\it Model building implications.}
The supersymmetry breaking mechanism
and the supersymmetry  breaking scale define to a large extent the phenomenology of supersymmetry.
Let us consider a breaking concisely described by
\be
\langle X \rangle = M + \theta^2 f \,. 
\ee
The breaking of supersymmetry has to be transmitted
from a BREAK sector to the SM (visible sector) through a messenger sector. 
If the messenger sector interactions are of
gravitational strength,
the intrinsic scale of supersymmetry breaking is of  order
$\sqrt{f} \gtrsim 10^{11}$ GeV, with MSSM sparicles having mass $\sim f/M$ with $M=M_P$. 
 It is  nevertheless likely that the usual gauge interactions of the SM play some role in the messenger sector and the messenger scale, $M$, might be anywhere between the Planck and the electroweak scale.
In gauge-mediated supersymmetry breaking (GMSB) it is SM gauge forces that transmit the supersymmetry breaking to the (MS)SM \cite{Dine:1981za,Dimopoulos:1981au,Dine:1981gu,Alvarez-Gaume:1981abe,Nappi:1982hm}.
A low-energy effective superpotential that captures the GMSB effect is the following
\be
W_{\rm med} = \lambda_{ij} X Q^i \overline Q^j \,,   
\ee
where $M$ can be read from above as the regulator of  the mass of the messenger fields $Q$ and $\bar{Q}$, and  $\lambda_{ij}$ are Yukawa couplings. 
The latter couplings  also play a role in the cosmological selection of the non-supersymetric metastable vacuum \cite{Dalianis:2010pq}. 
In the GSMB case the supersymmetry is broken at an intermediate scale, $f/M$, 
that characterizes the soft masses and the gravitino can be the LSP and hence a dark matter particle. 
The fact that the soft masses persist in the global SUSY limit since the BREAK sector does not decouple brings a contradiction with the GFL bound. 
Direct GMSB models often involve fully calculable SUSY-breaking sectors, with prominent examples the ISS \cite{Intriligator:2006dd}, the ITIY model \cite{Intriligator:1996pu,Izawa:1996pk},  as well as the canonical O’Raifeartaigh model \cite{ORaifeartaigh:1975nky}.

Apart from gauge and gravity mediation contributions there exists an ubiquitous form of mediation, known as ``anomaly mediation'' \cite{Randall:1998uk,Giudice:1998xp}.
 It induces a generic contribution to the gaugino masses and the A-terms, known as conformal anomaly mediated contribution. Such a contribution is dominant in models where there are no singlet fields present in the BREAK sector.
Anomaly mediation  involves three different anomalies: a super-Weyl anomaly, a K\"ahler anomaly, and a sigma-model anomaly \cite{Dine:2007me,DEramo:2012vvz}.
Remarkably SUGRA is not always a necessary ingredient as a version of anomaly mediation  appears even in the $M_P \rightarrow \infty$ limit  that corresponds to the sigma-model anomaly \cite{Dine:2007me}. K\"ahler mediation simply arises from linear couplings of SUSY-breaking fields in the K\"ahler potential, and thus appears in both global and local SUSY.
On the same footing with the gauge mediation scheme the aforementioned aspects of anomaly mediation are in contradiction with the GFL bound.  
An exception  
could be a scenario where 
supersymmetry breaking is gravitationally triggered and, hence, supersymmetry is restored as $M_{\rm Pl} \rightarrow \infty$, so that $V \rightarrow  0$ in the BREAK sector. This behaviour can be found  e.g. in no-scale supergravity structures \cite{KOUNNAS,Ellis:1983sf, Lahanas:1986uc}. In such a case the FL bound is respected in all sectors regardless the mediation scheme involved.

In general, only mediation schemes in which supersymmetry in the SM sector is restored in the global limit are FL favored, (with the only exception that of supersymmetric restoration in the BREAK sector in the global limit, as previously mentioned). 
The standard gravity mediation scheme is the  most notable and obvious example, where the mass spliting of sparticles vanishes when $M_{\rm Pl} \rightarrow \infty$. Moreover, an aspect of anomaly mediation where soft masses are proportional to the gravitino mass, also dubbed gravitino mediation,  is in accordance with the GFL bound. Therefore the GFL bound, although  insensitive to the energy scale of SUSY breaking, selects specific mediation schemes and has direct phenomenological implications.

{\it Phenomenological implications.}  Let us partly outline some basic  
  characteristics of the rich SUGRA phenomenology, see e.g. \cite{Aitchison:2007fn,Drees:2004jm,Binetruy:2012uov}. 
  In models of gravity-mediated supersymmetry breaking, gravity is the messenger of supersymmetry 
 and soft-supersymmetry-breaking parameters 
 arise as model-dependent multiples of the gravitino mass  $m_{3/2}$.\footnote{Bounds on the gravitino mass also arise from other swampland criteria \cite{Cribiori:2020use, Cribiori:2021gbf, Castellano:2021yye, DallAgata:2021nnr}, 
 especially it is argued in \cite{Anchordoqui:2023oqm} that gravity mediation will require a rather large supersymmetry breaking scale and a heavy gravitino.} 
Though gravity is flavor-blind the  supergravity mediation does not secure the flavor alignment of the soft masses. 
The experimental upper bound on flavor changing neutral currents (FCNC) amplitudes 
appeal either 
to the presence of extra symmetries, e.g. on the K\"ahler potential that suppress FCNC processes, or to very large sfermion masses that have to  lie above $10^3$ TeV \cite{Gabbiani:1996hi}.
In the former case the sparticles and particularly the gauginos which are expected to be lighter due to symmetry reasons can be within the reach of LHC. In the latter case, 
the scenario of phenomelogical interest is that of a split spectrum \cite{Arkani-Hamed:2004ymt,Giudice:2004tc} 
where scalar masses are dominated by gravity-mediated contributions at order $m_{3/2} \sim 10^4$ TeV, 
while gaugino masses are protected by an $R$-symmetry and arise at one loop via anomaly mediation at order $m_{1/2} \sim 10^2$ TeV. 
Although these sparticles are out of the LHC reach, gauge coupling unification  \cite{Dimopoulos:1981yj} favors a $\mu$-term that is much lighter than the scalars and the higgsinos may be produced at the LHC \cite{Arvanitaki:2012ps}. 

From the cosmological perspective the major implications of the GFL bound are, firstly, that
within the MSSM framework the LSP is the neutralino and contributes to the dark matter density in galaxies if $R$-parity is conserved and, secondly, 
that  the gravitino is unstable. 
The exact sparticle mass scale is unknown.
Gravitinos are produced via thermal scatterings, non-thermal decays of sfermions and possible decays of scalars beyond MSSM  such as the supersymmetry breaking field or other moduli. 
Focusing on the MSSM sector the gravitinos dominate the universe either for rather large enough reheating temperature 
or large enough sfermion masses.
The gravitinos decay at the temperature $T^\text{dec}_{3/2} \sim 7\, \text{MeV}\left({m_{3/2}}/{10^2\text{TeV}}\right)^{3/2}$. 
Apparently, a heavy gravitino  $m_{3/2}\gtrsim 10^2$ TeV avoids BBN complications and this is very welcome.
The gravitino decay also populates the universe with neutralinos. 
If gravitinos are heavy enough, $m_{3/2}\gg 10^4$ TeV, they decay promptly and $T^\text{dec}_{3/2}$ surpasses the neutralino freeze-out temperature
thus neutralinos reach a thermal equilibrium. 
For lighter gravitinos, the neutralinos produced by the graviton decay attain a non-thermal abundance with a value that depends critically on the gravitino number density. 
The requirement that neutralinos are not overproduced from decays and thermal scatterings 
places interesting constraints on both the sparticle spectrum, see e.g. \cite{Arkani-Hamed:2004zhs}, and the early universe cosmological scenarios.

{\it Summary.} 
As the motivation for supersymmetry shifts from low energy phenomenological considerations to 
motivations from UV completions of quantum gravity, 
it is natural to incorporate possible conditions that come from such a UV completion. 
Here we have taken a step in such direction aiming to restrict the possible SUSY breaking mediation scenaria. 
We have seen that the global limit of the FL bound of the swampland program disfavors mediation schemes that do not decouple together with gravity. 
The gauge mediation of SUSY breaking is one of these disfavored scenaria and therefore one could say that it is in the swampland; in other words it cannot arise within string theory, 
unless supersymmetry gets restored also in the breaking sector in the formal gravity decoupling limit (e.g. by breaking SUSY with no-scale models). 
This means in particular that, 
if the global limit interpretation of the FL bound is correct, 
it should not be possible to find within string theory models of gauge mediation where the breaking of supersymmetry is dynamical \cite{Izawa:1996pk,Intriligator:1996pu,Intriligator:2006dd}; 
this excludes large classes of GMSB scenaria \cite{Giudice:1998bp}. 
(Vice versa, if a fully fledged GMSB scenario with dynamical SUSY breaking is constructed and established within string theory then it will explicitly violate the GFL bound and point towards refinements.) 
The same conclusion can be drawn for particular anomaly mediation schemes in which visible sector fields have linear couplings to supersymmetry breaking fields in the K\"ahler potential.
Instead, gravity mediation (or variations of it) is favored because the mediation is decoupled when gravity decouples.

\section*{Acknowledgments}

The work of I.D. is supported by the Hellenic Foundation for Research and Innovation (H.F.R.I.) under the ``First Call for H.F.R.I. Research Projects to support Faculty members and Researchers and the procurement of high-cost research equipment grant'' (Project Number: 824). The work of F.F. is supported by the MIUR-PRIN contract 2017CC72MK003.

\end{document}